\documentstyle[12pt]{article}

\global\arraycolsep=1pt
\begin{document}
\baselineskip 18pt
\begin{titlepage}

\hfill CERN-TH/97-212

\hfill  IFUM-578/FT

\hfill {\tt hep-th/9708131}

\begin{center}
\hfill
\vskip .8in
{\large\bf On the Compton Scattering in String Theory}

\end{center}
\vskip .6in
\begin{center}
{\large Andrea Pasquinucci~$^{1,2}$ and Michela Petrini~$^1$}

\vskip .4in

{\sl $^1$ Dipartimento di Fisica, Universit\`a di Milano 

and INFN, Sezione di Milano 

via Celoria 16, 20133 Milan, Italy}

\vskip .2in

{\sl $^2$ Theory Division, CERN,
1211 Geneva 23, Switzerland}

\end{center}

\vskip1.2in

\begin{quotation}
\noindent {\bf Abstract}:
We explicitly compute the Compton amplitude for the scattering of a photon
and a (massless) ``electron/positron'' at tree level and one loop, 
in a four-dimensional fermionic heterotic string model. We comment on the 
relationship between the amplitudes we compute in string theory 
and the corresponding ones in field theory.

\end{quotation} 
\vspace{4pt}
\vfill

\noindent CERN-TH/97-212

\noindent August, 1997
\end{titlepage}
\eject

\newcommand{\be}{\begin{equation}}
\newcommand{\ee}{\end{equation}}
\newcommand{\ba}{\begin{eqnarray}}
\newcommand{\ea}{\end{eqnarray}}
\newcommand{\ban}{\begin{eqnarray*}}
\newcommand{\ean}{\end{eqnarray*}}
\newcommand{\brr}{\begin{array}}
\newcommand{\err}{\end{array}}
\newcommand{\bc}{\begin{center}}         
\newcommand{\ec}{\end{center}}
\newcommand{\sss}{\scriptscriptstyle}
\newcommand{\bea}{\begin{eqnarray}}
\newcommand{\eea}{\end{eqnarray}}
\newcommand{\bean}{\begin{eqnarray*}}
\newcommand{\eean}{\end{eqnarray*}}
\def\bbf#1{\mbox{\boldmath $#1$}}
\def\ds#1{\ooalign{$\hfil/\hfil$\crcr$#1$}}
\section{Introduction}

The computation of scattering amplitudes of string states through 
the Polyakov formula is one of the most powerful
tools we have to study the properties of first-quantized (perturbative)
string theories.  
Computations of tree-level (i.e.\ genus-zero) scattering amplitudes
in closed string theories often appeared in the literature, mostly
in the case where the external states are space-time bosonic particles. 
One-loop, i.e.\ genus-one, amplitudes have been computed in the case where
the external states are space-time bosons. These computations led to many
interesting results, in both string theory and field theory.
Indeed one can compare directly the scattering amplitudes for the
``same'' external states in field theory and string theory, or
one can take the field-theory limit of a string amplitude and 
compare it with the field theoretical one. Obviously the two expressions 
obtained in this way must be identical, but the way in which
string theory reproduces the field theoretical amplitudes can lead to
the discovery of new features of field theory (see for instance refs.\
\cite{Kap,BK,DiV}).

Very few one-loop amplitudes having space-time fermions as external states 
have appeared in the literature (see for example ref.\ \cite{At}), mostly
because of some technical issues appearing in the explicit 
computations of these amplitudes, as discussed for example
in ref.\ \cite{PR1}.

In this paper we present one of the simplest four-point 
one-loop scattering amplitude, which
involves external space-time fermions, that is the Compton scattering
of an ``electron/positron''  and a photon. Here we call ``electron''
(or ``positron'') a massless space-time fermion charged under a $U(1)$ 
component of the total gauge group. This state does not correspond 
directly to what we would usually call a, let us say, electron, since it is
massless, charged under other components of the total gauge
group, and also carries some (stringy) family labels. One can easily extend 
the results we present in this letter to the case of the scattering of 
a massless ``quark'' on a gluon, 
since this requires only some simple modifications
of the left-moving part of our equations.

In doing these computations, we have chosen a specific ``simple'' 
four-dimensional heterotic string model. 
One of the particular properties of this model is that its 
space-time spectrum depends on a set of parameters  and, 
as described in ref.\ \cite{PR1}, only for
some values of these parameters is supersymmetric.
Of course the string scattering amplitudes depend on the 
particular spectrum and on the details of the 
compactification from ten dimensions of the chosen model. 
Anyway, as experience has shown, we expect that the general features of the 
string scattering amplitudes are independent of the specific string
model chosen for the computation, in particular when one takes the 
field-theory limit. 


We consider a fermionic four-dimensional heterotic string model in 
the formalism of Kawai, Lewellen and Tye~\cite{KLT} 
(see also~\cite{Anto,Bluhm}). 
In this construction all the degrees of freedom
other than those associated with the 
four-dimensional space-time coordinates are described by 
free world-sheet fermions. The fields of the model are: 
four space-time coordinate fields $X^\mu(z, \bar{z})$,
twenty-two left-moving complex fermions
$\bar{\psi}_{(\bar{l})}(\bar{z})$ (with 
$\bar{l} = \bar{1}, \ldots , \overline{22}$), 
eleven right-moving complex fermions
$\psi_{(l)}(z)$ (with $l= 23, \ldots,33$) of which $\psi_{(32)}$ and 
$\psi_{(33)}$ are the world-sheet superpartners of the space-time coordinates
$X_\mu$, right-moving superghosts 
$\beta, \gamma$,
left- and right-moving reparametrization ghosts 
$\bar{b}, \bar{c}$ and $b,c$.
In ref.~\cite{PR1} the reader can find a complete description of
the model, and further details on the conventions 
on the spin structures, and the vertex operators of the model 
under consideration can be found in refs.~\cite{PR2,PR3}.
The gauge group of the model is
$SO(14)\otimes SO(14) \otimes SO(4) \otimes U(1) \otimes SO(10)$.
The $U(1)$ charge, which we will be interested in, is associated 
with the seventeenth left-moving fermion.
The vertex operator associated with the $U(1)$ gauge boson (the photon) 
with polarization $\epsilon$ 
and momentum $k$ in the superghost picture $q=-1$ is~\cite{FMS}
\be
{\cal V}^{(-1)}_{{\rm photon}} (z,\bar{z};k;\epsilon) = {\kappa 
\over \pi} \bar\psi_{(\overline{17})}\bar\psi_{(\overline{17})}^* (\bar{z}) \
\epsilon \cdot \psi (z) \, e^{-\phi(z)} \, (c_{(34)})^{-1} e^{ik\cdot
X(z,\bar{z})},
\ee
where the gravitational coupling $\kappa$ is related to Newton's constant by 
$\kappa^2 = 8\pi G_N$, $\epsilon \cdot \epsilon = 1$, 
$k^2 = \epsilon \cdot k = 0$
and $c_{(34)}$ is the cocycle associated to the superghosts. 
The picture-changed version of the same vertex is
\be
{\cal V}^{(0)}_{{\rm photon}} (z,\bar{z};k;\epsilon) = 
-i {\kappa \over \pi} \bar\psi_{(
\overline{17})}\bar\psi_{(\overline{17})}^* (\bar{z}) \
\left[ 
\epsilon \cdot \partial_z X(z) - i k\cdot \psi (z) \epsilon \cdot \psi (z) 
\right] \,  e^{ik\cdot X(z,\bar{z})} \ .
\ee
The vertex operator for the massless fermions of momentum $k$, $U(1)$ charge 
$\pm 1/2$ and superghost charge $q= -1/2$ is given by
\be
{\cal V}^{(-1/2)} (z,\bar{z};k;{\bf V})= 
 \frac{\kappa}{\pi} \ {\bf V}^A 
S_A (z,\bar{z}) \
e^{-\frac12\phi(z)} (c_{(34)})^{-1/2}\ e^{ik\cdot X(z,\bar{z})}, 
\ee
where $S_A$ is the spin field that creates the Ramond ground state 
representing the fermion from the conformal vacuum: 
\be
S_A \ = \ 
\left(\prod_{l=1}^7 \bar{S}^{(\bar{l})}_{\bar{a}_l} (\bar{z})\right) \ 
\bar{S}^{(\bar{17})}_{\bar{a}_{17}} (\bar{z})
\left(\prod_{l=24,28,29} S^{(l)}_{a_l}(z)\right) \  S^{(l)}_{\alpha}(z)
\ee
and the ``spinor'' ${\bf V}^A$ decomposes accordingly.
All indices $(\bar{a}_l,a_l)$ take values $\pm 1/2$: 
$(\bar{a}_1,\ldots,\bar{a}_7)$ are indices 
of the first $SO(14)$; 
$\bar{a}_{17} = \pm 1/2$ is the $U(1)$ charge;
$\alpha= (a_{32}, a_{33})$
is the four-dimensional space-time spinor index, while 
$(a_{24},a_{28},a_{29})$ are just enumerative family indices.
The explicit expression of the spin fields is obtained by bosonizing the 
world-sheet fermions, see refs.~\cite{FMS,PR1,PR2}. 
We have introduced the dimensionless momentum 
$k_\mu=\sqrt{\frac{\alpha'}{2}}p_\mu$ and
the Minkowski space-time metric is $\eta_{\mu\nu}=(-1,1,1,1)$.

\section{The tree-level amplitude}

We start by discussing the tree-level scattering and comparing the
string-theory amplitude with the field-theory one. The tree-level,
i.e.\ genus-zero, string Compton scattering amplitude turns out
to be practically independent of the chosen string model, as 
will be obvious from eq.\ (\ref{treelev2}).

We adopt the notations of ref.~\cite{PR3} for the general formula 
expressing the connected part of a string scattering amplitude as 
a two-dimensional correlation function of vertex operators.
At tree level, the amplitude for the scattering of two photons and two massles
fermions is then written as
\bea
T_{g=0}& &(\ e^{\pm}\ + \ \gamma \ \rightarrow \ e^{\pm} \ + 
\ \gamma ) \ = \nonumber\\
  & &  C_{g=0} \int{\rm d}^2z_1\, \langle \Pi(w)
{\cal V}^{(-1)}_{\rm photon} (z_1,\bar{z}_1;k_1;\epsilon_1)
\bar{c}(\bar{z}_2)c(z_2)
{\cal V}^{(-1)}_{\rm photon} (z_2,\bar{z}_2;k_2;\epsilon_2) \nonumber\\
  & & \bar{c}(\bar{z}_3)c(z_3){\cal V}^{(-1/2)} (z_3,\bar{z}_3;k_3;{\bf W}_3)
\bar{c}(\bar{z}_4)c(z_4) 
{\cal V}^{(-1/2)}(z_4,\bar{z}_4;k_4;{\bf V}_4)\rangle\ , \label{treelev1}
\eea
where $C_{g=0}=-4\pi^3/(\kappa^2\alpha')$ is a constant giving the proper
normalization of the vacuum amplitude~\cite{Kaj, PR3}, ${\bf V}_4$ is 
the ``spinor" of
the incoming electron, while the vector ${\bf W}_3$ is associated with the
outgoing one as in eq.\ (8.11) of ref.~\cite{PR3}, see also below.

At tree level one generically takes all space-time fermionic vertex 
operators to have the superghost charge $q=-1/2$  and all the bosonic 
ones to have the superghost charge $q=-1$. We denote by
$\Pi(w)$ the Picture-Changing Operator (PCO) needed to compensate the
superghost vacuum charge~\cite{DPh, FMS}. 
The amplitude does not depend on the PCO 
insertion point $w$, even if, in general, it is not so straightforward
to show such a property on the final form of the amplitude. 
In the case at hand, however, it is easy to show that 
in the final form of the amplitude the dependence on the
PCO insertion point  $w$ cancels explicitly.
Of course, one can always take the limit $w\rightarrow z_1$
(or $w\rightarrow z_2$) in eq.\ (\ref{treelev1}), 
eliminating the PCO operator from the beginning
and transforming one photon vertex 
operator into its picture-changed version 
with superghost charge $q=0$.  Technically it is often
convenient to keep the PCO at a generic point on the world-sheet, since
the independence of the final result from it is a
valuable check on the correctness of the computation.\par
After standard manipulations, the final form of the tree amplitude is
\bea
T_{g=0} &(&e^{\pm} + \gamma \ \rightarrow \ e^{\pm} + \gamma)= \nonumber\\
  & - & \frac{\kappa^2}{(\alpha')^{3/2}}  
\left[ \left(\frac{1}{s} +\frac{1}{u} \right) 
{\bf W}^T_3 \ds{p_1}\ds{\epsilon_1}\ds{\epsilon_2}
{\bf C V}_4 - \left(\frac{2}{s} \right) (\epsilon_1\epsilon_2) 
{\bf W}^T_3 \ds{p_1}{\bf C V}_4  \right. \nonumber\\
 & &  \qquad\quad+\left( \frac{2}{s} \right) 
(\epsilon_2 p_1) {\bf W}^T_3 \ds{\epsilon_1}{\bf C V}_4 
+ \left. \left( \frac{2}{s} ( \epsilon_1 p_4)
 - \frac{2}{u} (\epsilon_1 p_3) \right)
{\bf W}^T_3 \ds{\epsilon_2}{\bf C V}_4 \right] \nonumber\\ 
 &\times & \left[ 1- \frac{\alpha'su}{t(+\frac14\alpha't)} \right]
\frac{\Gamma\left(1-\frac14\alpha' s \right)
\Gamma\left(1-\frac14\alpha' t \right)
\Gamma\left(1-\frac14\alpha' u \right)}
{\Gamma\left(1+\frac14\alpha' s \right)
\Gamma\left(1+\frac14\alpha' t \right)
\Gamma\left(1+\frac14\alpha' u \right)} \label{treelev2}\ ,
\eea
where the Mandelstam variables $s,t,u$ are defined as: 
$s = - (p_1 + p_4)^2 = - (p_2 + p_3)^2$, 
$t = - (p_1 + p_2)^2 = - (p_3 + p_4)^2$, 
$u  = - (p_2 + p_4)^2 = - (p_1 + p_3)^2$.

The ``spinors'' {\bf V} are normalized according to ref.~\cite{PR3}, i.e.\
${\bf V}^{\dagger}(p){\bf V}(p)= \sqrt{\alpha'}|p_0|$,
so that there is a factor $(\alpha')^{1/4}/\sqrt{2}$ with respect to
the normalization of spinors in field theory,\footnotemark 
\footnotetext{~For our conventions in field theory and the 
normalization of spinors we follow, for instance, ref.\ \cite{LeB}, 
with the opposite Minkowski metric.} 
and the vector ${\bf W}^T_3$ is related to the ``spinor'' 
${\bf V}^T_3$, describing the outgoing electron, by 
${\bf W}^T_3={\bf V}^T_3 \sigma_1^{(34)}\Sigma$ (see ref.\ \cite{PR3} for
more details).
Finally, the gauge coupling constant is expressed in terms of the 
constant $\kappa$ by the relation
$e^2=\kappa^2/(2 \alpha')$. 

After these substitutions have been performed, the string amplitude can be
easily compared with the field theoretical one. 
It is interesting to note that, after expressing eq.~(\ref{treelev2}) in
terms of the field-theory variables, the dependence on
$\alpha'$ is limited to the last line of eq.~(\ref{treelev2}). 
Indeed the first two lines of eq.~(\ref{treelev2}) give exactly the 
tree-level field theory result for the
Compton amplitude (see for instance~\cite{Fey, LeB}):
\bea
T(&e^{\pm}&  + \gamma \ \rightarrow \ e^{\pm} + \gamma)= -   e^2  
\left[ \left(\frac{1}{s} +\frac{1}{u}\right) 
{\bar{\bf u}}(p_3) \ds{p_1}\ds{\epsilon_1}\ds{\epsilon_2}{\bf u}(p_4) 
- \left(\frac{2}{s} \right) (\epsilon_1\epsilon_2)
{\bar{\bf u}}(p_3) \ds{p_1} {\bf u}(p_4)  \right. \nonumber\\
 & & \qquad +\left( \frac{2}{s} \right) 
(\epsilon_2 p_1){\bar{\bf u}}(p_3)\ds{\epsilon_1}{\bf u}(p_4) 
+ \left. \left( \frac{2}{s}  (\epsilon_1 p_4) 
 - \frac{2}{u} (\epsilon_1 p_3) \right) 
{\bar{\bf u}}(p_3) \ds{\epsilon_2}{\bf u}(p_4) \right] \label{treelevft}
\eea
and in the $\alpha' \rightarrow 0$ limit the last line of
eq.~(\ref{treelev2}) goes exactly to $+1$. 

Thus, at tree level, the string Compton amplitude has the interesting feature
that it factorizes in a part that coincides with the field-theory
amplitude and a part that gives the string-theory corrections to the 
scattering. The stringy correction is in the usual Veneziano form, showing
the appearance of the poles corresponding to all possible intermediate 
string states.

\section{The one-loop amplitude}

In this section we present the one-loop, i.e.\ genus-one, Compton
scattering amplitude in our four-dimensional string model. 
We have chosen to compute the scattering of a chiral massless fermion
on a photon since only string massless states survive in the 
field-theory limit, which allows us to make easy comparisons 
with the field-theory results. 

The presence of massless external states implies that our amplitude suffers
from infrared divergences which are the same as the ones appearing in field
theory and are due to the emission/absorption of soft photons/electrons. 
The easiest way to deal with these divergences is to 
introduce an appropriate infrared cut-off. 

The string scattering amplitude is free from the ultraviolet divergences 
that appear in field theory and it automatically regulates also the
chiral massless fermions describing our electrons. These divergences 
of course reappear in the field-theory limit basically as divergences 
in $\alpha'\rightarrow 0$. The string scattering amplitude is not anyway
free of divergences; instead, as is well known, there appear divergences
in the integrations over the moduli. 
The physical interpretation of these divergences has been discussed
for example in refs.\ 
\cite{PR3}, \cite{Wein}---\cite{PR5}, 
and is related to the unitarity
of the scattering amplitude. The amplitude that we present below is 
unregulated, which means formally Hermitian, and to obtain the amplitude
with the poles and cuts required by unitarity one has to apply a procedure
of analytic continuation (the stringy version of the 
Feynman ``$+i\epsilon$''), 
similar for example to one of those described in refs.\ \cite{DP2,Bere,Mont}. 

Keeping these points in mind, we could try to compare the 
string scattering amplitude that we will present in this section,
with the corresponding one in field theory, 
as we did for the tree-level computation in the previous section. 
In this case, the integrals over the moduli in the string amplitude
correspond to the integrals over the Schwinger parameters in
field theory (see for example ref.\ \cite{BK}). 
At a first look,\footnotemark 
\footnotetext{~We do not display the field-theory expression of the
scattering amplitude at one loop, but the reader can easily derive it
her/himself.} one immediately notices that the string theory expression
we give below contains more kinematical terms than the one in field theory. 
Of course, one should compare the
string- and field-theory expressions only after having done the integrals
over the Schwinger parameters in field theory, and the integrals over
the moduli and the sum over the spin-structures in string theory. In field
theory the integrals can be done, but in string theory the integrals
and the sums usually cannot. This prevents us from making any 
particular claim about the properties of the string-theory amplitude before
making a suitable approximation or taking the field-theory limit. 


The starting point of the computation of the one-loop Compton
scattering amplitude in string theory is given by 
the connected part of the S matrix:
\bea
T_{g=1}&(& e^{\pm} \ + \ \gamma  \ \rightarrow \ e^{\pm} \ 
+ \ \gamma )= \nonumber\\
   &C&_{g=1} \int {\rm d}^2\tau {\rm d}^2z_2 {\rm d}^2z_3 
      {\rm d}^2z_4 \
      \sum_{m_i,n_j} C^{\bf\alpha}_{\bf\beta} \
      \langle \langle \vert (\eta_{\tau}\vert b) (\eta_{z_2}\vert b) 
      (\eta_{z_3}\vert b)(\eta_{z_4}\vert b) \nonumber\\
  &\times&  c(z_1)c(z_2)c(z_3)c(z_4)\vert^2 \
      \Pi(w) \ {\cal V}^{(0)}_{\rm photon} 
      (z_1,\bar{z}_1;k_1;\epsilon_1)
      {\cal V}^{(0)}_{\rm photon} (z_2,\bar{z}_2;k_2;\epsilon_2)\nonumber\\
  &\times & {\cal V}^{(-1/2)} (z_3,\bar{z}_3;k_3;{\bf W}_3) \ 
      {\cal V}^{(-1/2)} (z_4,\bar{z}_4;k_4;{\bf V}_4) \rangle \rangle, 
\label{oneloop1}
\eea
where $C_{g=1}=1/(2 \pi \alpha')^2$ ~\cite{Kaj, PR3}.
In higher-genus computations, it is convenient \cite{PR1,PR3}
to use bosonic vertex
operators in the zero super ghost picture and fermionic vertex operators
in the $-1/2$ picture. In our case this leaves only one PCO at an
arbitrary point $w$. As a consistency check we have verified
that the final expression of the amplitude, eq.\ (\ref{oneloop2}), 
is independent of $w$.
Notice that, as was already the case in ref.\ \cite{PR1}, it is not possible 
to explicitly eliminate  $w$ from eq.\ (\ref{oneloop2}), as we did
at tree level, but since this equation does not depend on $w$, 
one can fix $w$ to the most convenient value. 

The technical details of the computation of eq.\ (\ref{oneloop1}),
which has been done following the procedure of ref.\ \cite{PR1}, will be
discussed elsewhere. Let us only note that the main difficulties
come from the presence of the spin fields. One way of computing
these correlators is to bosonize all world-sheet fermions~\cite{FMS,Koste}. 
At this point the computation of the correlators becomes almost 
trivial, but Lorentz covariance is lost and can be recovered only after
some technical steps that involve, among other things, also the use
of some non-trivial identities in theta functions, mostly of the
form of the trisecantic identity \cite{Fay}.

The final form of the one-loop amplitude is
\bea
T_{g=1} (\ e^{\pm} \ + \ \gamma & & \ \rightarrow \ e^{\pm} \ 
+ \ \gamma )= \nonumber\\
     & &  \frac{e^4}{\pi^6}
\sum_{m_i,n_j} C^{\alpha}_{\beta} \ e^{2\pi i K_{GSO} }
\int {{\rm d}^2 \tau\over ({\rm Im}\tau)^2 } \ 
(\bar{\eta} (\bar{\tau}) )^{-24} (\eta(\tau))^{-12} \nonumber\\
     & & \times\int {\rm d}^2z_2 {\rm d}^2z_3 {\rm d}^2z_4 \ 
\frac{\sqrt{\omega(z_3)\omega(z_4)}}
{\bar{\omega} (\bar{z}_1) \omega(z_1)\omega (w)} \
exp\left[\sum_{i<j}(k_ik_j) G_B (z_i,z_j) \right] \nonumber\\
 & & \times{\cal T}_L \left[{}^{\bar{\alpha}}_{\bar{\beta}} \right]
(z_1,z_2,z_3,z_4,w)   
\times {\cal T}_R \left[{}^{\alpha}_{\beta} \right]
(z_1,z_2,z_3,z_4,w)\ , \label{oneloop2}
\eea
where $\omega(z)=1/z$ and
\bea
{\cal T}_L & (& z_1,z_2,z_3,z_4,w) =
\prod_{l=1-7,17} \bar{\Theta}
\left[{}^{\bar{\alpha}_{l}}_{\bar{\beta}_{l}} \right] 
(\frac12 \bar{\nu}_{34} \vert \bar{\tau} )\times
\prod_{l=8}^{16} \prod_{l=18}^{22}\bar{\Theta} \left[
{}^{\bar{\alpha}_{l}}_{\bar{\beta}_{l}} \right] ( 0 \vert \bar{\tau} )
\times (\bar{E} (\bar{z}_3,\bar{z}_4))^{-2} \nonumber\\
                 & \times & \left\{ \partial_{\bar{z}_1}
\partial_{\bar{z}_2} \log \bar{E}(\bar{z}_1,\bar{z}_2) 
+\frac14 \ \partial_{\bar{z}_1}\log \frac{\bar{E}(\bar{z}_1,\bar{z}_3)}
{\bar{E}(\bar{z}_1,\bar{z}_4)} \ 
 \partial_{\bar{z}_2}\log \frac{\bar{E}(\bar{z}_2,\bar{z}_3)}
{\bar{E}(\bar{z}_2,\bar{z}_4)} \right. \nonumber\\
                 &  & \quad+ \frac12 \frac{\bar{\omega}(\bar{z}_1)}{2\pi i} \  
\partial_{\bar{z}_2}\log \frac{\bar{E}(\bar{z}_2,\bar{z}_3)}
{\bar{E}(\bar{z}_2,\bar{z}_4)}\partial_{\nu} \log \bar{\Theta}
\left[{}^{\bar{\alpha}_{17}}_{\bar{\beta}_{17}} \right] 
(\nu\vert\bar{\tau})\vert_{\nu=\frac12\bar{\nu}_{34}} \nonumber\\
                 &  & \quad + \frac12
\frac{\bar{\omega}(\bar{z}_2)}{2\pi i} \  
\partial_{\bar{z}_1}\log \frac{\bar{E}(\bar{z}_1,\bar{z}_3)}
{\bar{E}(\bar{z}_1,\bar{z}_4)} \partial_{\nu} \log \bar{\Theta}
\left[{}^{\bar{\alpha}_{17}}_{\bar{\beta}_{17}} \right] 
(\nu\vert\bar{\tau})\vert_{\nu=\frac12\bar{\nu}_{34}} \nonumber\\
                 &  & \quad + \left. \frac{\bar{\omega}(\bar{z}_1)}{2\pi i}
\frac{\bar{\omega}(\bar{z}_2)}{2\pi i} \ 
\left(\bar{\Theta} \left[{}^{\bar{\alpha}_{17}}_{\bar{\beta}_{17}} \right]
(\frac{\scriptstyle 1}{\scriptstyle 2}\bar{\nu}_{34}\vert\bar{\tau}) 
\right)^{-1} \ \partial^2_{\nu} 
\bar{\Theta}
(\nu\vert\bar{\tau})\vert_{\nu=\frac12\bar{\nu}_{34}} \right\}, 
\eea
\bea
{\cal T}_R  & (  & z_1,z_2,z_3,z_4,w) = 
\prod_{l=24,28,29,32}\Theta
\left[{}^{\alpha_{l}}_{\beta_{l}} \right] (
\frac{\scriptstyle 1}{\scriptstyle 2}\nu_{34} \vert \tau )
\prod_{l=23,25,26,27,30,31}\Theta
\left[{}^{\alpha_{l}}_{\beta_{l}} \right] (0 \vert \tau)\nonumber\\
                 &\times&  \frac{(-1)^{S_{33}}}{\sqrt2} (E(z_3,z_4))^{-1} 
\left\{{\bf W}^T_3 \ds{k_1}\ds{\epsilon_1}\ds{\epsilon_2}
\left( \Gamma^5 \right)^{\tilde{S}}  
{\bf C V}_4  \ {\cal A}_1(z_1,z_2,z_3,z_4,w) \right. +\nonumber\\
                 &  & \quad + {\bf W}^T_3 \ds{k_1} 
\left( \Gamma^5 \right)^{\tilde{S}}
{\bf C V}_4  \ {\cal A}_2(z_1,z_2,z_3,z_4,w) + \nonumber\\
                 &  & \quad+ {\bf W}^T_3  \ds{\epsilon_1} 
\left( \Gamma^5 \right)^{\tilde{S}}
{\bf C V}_4 \ {\cal A}_3(z_1,z_2,z_3,z_4,w) + \nonumber\\
                 &  & \quad+ \left.  {\bf W}^T_3 \ds{\epsilon_2} 
\left( \Gamma^5 \right)^{\tilde{S}}
{\bf C V}_4 \ {\cal A}_4(z_1,z_2,z_3,z_4,w) \right\}\ ,
\eea
with $\nu_i=\frac1{2\pi i}\log z_i$ and $\nu_{ij}=\nu_i-\nu_j$. 
In eq.~(\ref{oneloop2}), $K_{GSO}$  is a phase factor that depends 
on the variables 
$k_{ij}$ parametrizing the GSO projection:
\bea
K_{GSO} & = & (k_{02}+k_{12}+k_{14}+k_{23}+k_{24}+k_{34}+ 1/2 )S_1 \nonumber\\
&&+ (k_{00}+k_{01}+k_{02}+k_{03}+k_{04}+k_{12}
+ k_{23}+k_{24}+ 1/2 )S_{17} \nonumber\\
&&+ (k_{00}+k_{01}+k_{03}+k_{04}+k_{13}+k_{34}+1/2 )S_{24} \nonumber\\
&&+ (k_{00}+k_{01}+k_{03}+k_{04}+k_{13}+k_{14}+1/2 )S_{29}\ ,
\eea
$S_i=(1-2\alpha_i)(1+2\beta_i)$ and 
$\tilde{S}=S_{17}+S_{24}+S_{29}+S_{33}$.
The coefficients ${\cal A}_i$ are functions of the external momenta, 
polarization vectors and world-sheet coordinates:\footnotemark 
\footnotetext{~Even if not explicitly written, all functions listed 
below depend also on the world-sheet coordinates $z_3,z_4$.
For our conventions on the theta functions and the prime form, see 
ref.~\cite{PR1}.}
\bean
{\cal A}_1(&z_1&,z_2,w) = (k_1k_2) \left[\partial_w G_B(w,z_1) 
- \partial_w G_B(w,z_2)\right]G^+(z_1,z_2)G^-(z_1,z_2) \\   
                     & + &  \frac12 \sum_{j=1}^4 (k_1k_j)
\partial_{z_1} G_B(z_j,z_1) \partial_w G_B(z_1,w) I(z_2) + \\
                     & + & \frac12 \sum_{j=1}^4 (k_2k_j)
\partial_{z_2}G_B(z_j,z_2)\partial_w G_B(z_2,w)I(z_1)  + \\    
                     & + & \sum_{j=1}^4 \partial_w G_B(w,z_j)
\left[(k_1k_j){\cal I}(z_2)G^+(z_1,w)
+(k_2k_j){\cal I}(z_1)G^+(z_2,w)\right]
\frac{G^-(z_1,z_2)}{{\cal I}(w)},
\eean
\bean
{\cal A}_2 ( & z_1, & z_2,w) = \left[\partial_w G_B(z_1,w) - 
\partial_w G_B(z_2,w) \right] 
\left[(\epsilon_1\epsilon_2)
\partial_{z_1}\partial_{z_2}G_B(z_1,z_2) + \right. \\
           &  & \quad + \sum_{j,i=1}^4 (\epsilon_1k_i)(\epsilon_2k_j)
\partial_{z_1}G_B(z_i,z_1) \partial_{z_2}G_B(z_j,z_2)] + \\
           & + & (\epsilon_1k_3) \sum_{j=1}^4(\epsilon_2k_j)  
\partial_{z_2}G_B(z_j,z_2)I(z_1) 
\left[\partial_w G_B(z_2,w) - \partial_w G_B(z_3,w)\right] + \\
           & - & (\epsilon_2k_3) \sum_{j=1}^4(\epsilon_1k_j)  
\partial_{z_1}G_B(z_j,z_1)I(z_2) \left[\partial_w G_B(z_1,w) - 
\partial_w G_B(z_3,w)\right] + \\
           & - & (\epsilon_1\epsilon_2)
\sum_{j=1}^4 \left[(k_2k_j) \partial_{z_2} G_B(z_j,z_2)\partial_w G_B(z_2,w)
B^+(z_1,w) + \right. \\
           &  & \quad - \left. (k_1k_j)
\partial_{z_1} G_B(z_j,z_1) \partial_w G_B(z_1,w)B^-(z_2,w) \right] + \\
           & - & \sum_{j,i=1}^4 (\epsilon_1k_i)(\epsilon_2k_j)
\left[\partial_{z_1} G_B(z_i,z_1)\partial_w G_B(z_j,w)B^+(z_2,w) + \right. \\
           &  & \quad - \left. \partial_{z_2} G_B(z_j,z_2)
\partial_w G_B(z_i,w)B^+(z_1,w) \right] + \\
           & + & \sum_{j=1}^4 \partial_w G_B(z_j,w)
\left\{ \left[(\epsilon_1k_2)(\epsilon_2k_j)-
(\epsilon_1\epsilon_2)(k_2k_j)\right]C_2(z_1,z_2,w) + \right. \\
           &  & \quad +\left. \left[(\epsilon_2k_1)(\epsilon_1k_j)-
(\epsilon_1\epsilon_2)(k_1k_j)\right] C_1(z_1,z_2,w) \right\} + \\
           & + & 2 \sum_{j=1}^4 \partial_w G_B(z_j,w)
\left[(\epsilon_1k_3)(\epsilon_2k_j)D_2(z_1,z_2,w) + 
(\epsilon_2k_4)(\epsilon_1k_j) D_1(z_1,z_2,w)\right] + \\
           & + & \left[(\epsilon_1k_2)(\epsilon_2k_1) -        
(\epsilon_1\epsilon_2)(k_1k_2)\right]
\left[\partial_w G_B(z_1,w) - \partial_w G_B(z_2,w)\right] \times  \\
           & & \quad \times \left[G^+(z_1,z_2)^2 
+2G^+(z_1,z_2)G^-(z_1,z_2)\right] +\\
           & + & 2  \ G^+(z_1,z_2)G^-(z_1,z_2) \{(\epsilon_1k_2) 
\sum_{j=1,3}(\epsilon_2k_j)\left[\partial_w G_B(z_2,w) - 
\partial_w G_B(z_j,w)\right] +  \\
           &   & \quad+\left[(\epsilon_1\epsilon_2)(k_1k_2)+
(\epsilon_1k_3)(\epsilon_2k_1)\right] \left[\partial_w G_B(z_1,w) - 
\partial_w G_B(z_3,w)\right]\},
\eean
\bean
{\cal A}_3(& z_1 &,z_2,w) = \partial_w G_B(z_1,w)[(\epsilon_2k_1)
\partial_{z_1} G_B(z_1,z_2)\sum_{j=1}^4(k_jk_2)\partial_{z_2}
G_B(z_2,z_j)+  \\
           &  & \quad - \sum_{j,i=1}^4 (k_1k_i)(\epsilon_2k_j)
\partial_{z_1}G_B(z_i,z_1)\partial_{z_2}G_B(z_j,z_2)] + \\
           & - & (k_1k_3)\sum_{j=1}^4 (\epsilon_2k_j)
\partial_{z_2} G_B(z_j,z_2) 
\left[\partial_w G_B(z_2,w) - \partial_w G_B(z_3,w)\right]I(z_1)+\\
           & - & (\epsilon_2k_4)\partial_w G_B(z_1,w)
\sum_{j=1}^4(k_1k_j)\partial_{z_1} G_B(z_j,z_1)I(z_2)+ \\
           & - & \sum_{j=1}^4 \left[(k_1k_2)(\epsilon_2k_j)-
(\epsilon_2k_1)(k_2k_j)\right] \partial_w G_B(z_j,w)C_2(z_1,z_2,w)+ \\
           & + & B^+(z_1,w)\sum_{j=1}^4 \partial_{z_2} G_B(z_j,z_2)
 \left[(k_2k_j) (\epsilon_2k_1)\partial_w G_B(z_2,w) + \right. \\
           &  & \quad - \sum_{i=1}^4(\epsilon_2k_j)(k_1k_i)
\partial_w G_B(z_i,w)] + \\
           & + & 2 G^+(z_1,z_2)G^-(z_1,z_2) \left\{(\epsilon_2k_1)(k_1k_3)
\left[\partial_w G_B(z_3,w) - \partial_w G_B(z_1,w)\right] + \right. \\
           &  & \quad - \left. (\epsilon_2k_4)(k_1k_2)
\left[\partial_w G_B(z_3,w) - \partial_w G_B(z_2,w)\right] \right\}+ \\
           & - & 2 \sum_{j=1}^4 \partial_w G_B(z_j,w)
\left[(\epsilon_2k_4)(k_1k_j)D_1(z_1,z_2,w)+
(k_1k_3)(\epsilon_2k_j)D_2(z_1,z_2,w)\right],
\eean
\bean
{\cal A}_4(& z_1 &, z_2,w)= \partial_w G_B(z_2,w)[(\epsilon_1k_2)
\partial_{z_2}G_B(z_1,z_2)\sum_{j=1}^4(k_jk_1)\partial_{z_1}
G_B(z_1,z_j)+ \\
           &  & \quad - \sum_{j,i=1}^4(k_2k_i)(\epsilon_1k_j)
\partial_{z_1}G_B(z_j,z_1)\partial_{z_2}G_B(z_i,z_2)] + \\
           & - & (k_2k_3)\sum_{j=1}^4 (\epsilon_1k_j)
\partial_{z_1} G_B(z_j,z_1)\left[\partial_w G_B(z_1,w) - 
\partial_w G_B(z_3,w)\right] I(z_2)+\\  
           & + & (\epsilon_1k_4)\partial_w G_B(z_1,w)
\sum_{j=1}^4(k_1k_j)\partial_{z_1} G_B(z_j,z_1)I(z_2)+ \\
           & + & \sum_{j=1}^4 \left[(k_1k_2)(\epsilon_1k_j)-
(\epsilon_1k_2)(k_1k_j)\right] \partial_w G_B(z_j,w)C_1(z_1,z_2,w)+ \\
           & - & B^+(z_2,w)\sum_{j=1}^4 \partial_{z_1} G_B(z_j,z_1) 
\left[\sum_{i=1}^4(\epsilon_1k_j)(k_2k_i)
\partial_w G_B(z_i,w) +\right. \\
           &   & \quad - (k_1k_j)(\epsilon_1k_2)
\partial_w G_B(z_1,w) \Big] \\
           & + & 2G^+(z_1,z_2)G^-(z_1,z_2)\left[(\epsilon_1k_3)(k_1k_2) -
(\epsilon_1k_2)(k_1k_3)\right]\times\\
           &   &\quad\times \left[\partial_w G_B(z_3,w) - 
\partial_w G_B(z_2,w)\right] +\\
           & + & 2 \sum_{j=1}^4 \partial_w G_B(z_j,w)  
\left[(\epsilon_1k_j)(k_1k_3) - 
(k_1k_j)(\epsilon_1k_3)\right]D_1(z_1,z_2,w),
\eean
and finally
$$
G_B(x,\bar{x};y, \bar{y})  = 
2 \left[ \log \vert
E(x,y)\vert - \frac12 {\rm Re}\left( \int_{y}^{x}\omega \right)^2 
\frac{1}{2\pi {\rm Im}\tau} \right], 
$$
$$
{\cal I}\left[{}^{\alpha}_{\beta} \right] (z)  = 
\sqrt{\frac{E(z_3,z_4)} {E(z,z_3)E(z,z_4)}} 
\frac{\Theta \left[{}^{\alpha}_{\beta} \right] (\nu_z-\frac12 \nu_3 -
\frac12 \nu_4\vert \tau)}
{\Theta \left[ {}^{\alpha}_{\beta} \right] 
(\frac12 \nu_{34} \vert\tau)}, 
$$
$$
I \left[{}^{\alpha}_{\beta} \right] (z) = 
\partial_z \log \frac{E(z,z_3)}{E(z,z_4)} 
+ 2 \frac{\omega(z)}{2\pi i} 
\partial_{\nu} \log {\Theta \left[{}^{\alpha}_{\beta} \right] (\nu \vert \tau)}
\vert_{\nu=\frac12\nu_{34}},  
$$
\bean
 G^{\pm}\left[{}^{\alpha}_{\beta} \right] (z,w) & = &  
\frac{1}{2 E(z,w)}
\left\{\frac{\Theta \left[{}^{\alpha}_{\beta} \right](\nu_{zw} + 
\frac12 \nu_{34}\vert\tau)}
{\Theta \left[{}^{\alpha}_{\beta} \right](\frac12\nu_{34}\vert\tau)}
\sqrt{\frac{E(z,z_3) E(w,z_4)}{E(w,z_3) E(z,z_4)}} \right. \\
 & &   \left. \pm \ \frac{\Theta \left[{}^{\alpha}_{\beta} \right]
(\nu_{wz} + \frac12 \nu_{34}\vert\tau)} 
{\Theta \left[{}^{\alpha}_{\beta} \right]
(\frac12\nu_{34}\vert\tau)} 
\sqrt{\frac{E(w,z_3) E(z,z_4)}{E(z,z_3) E(w,z_4)}} \right\}, 
\eean                     
$$
B^{\pm} \left[{}^{\alpha}_{\beta} \right] (z_i,w)=  
\frac{{\cal I}(z_i)}{{\cal I}(w)}
\left[G^+(z_i,w) \pm  G^-(z_i,w)\right], 
$$
$$
D_{1,2} \left[{}^{\alpha}_{\beta} \right] (z_1,z_2,w) = 
\frac{{\cal I}(z_{2,1})}{{\cal I}(w)}G^+(z_{1,2},w)G^-(z_1,z_2), 
$$
\bean
C_{1,2} \left[{}^{\alpha}_{\beta} \right](z_1,z_2,w) & = &
\left[G^+(z_1,z_2)G^-(z_1,z_2) + \right. \\
   & & + \left. \frac{{\cal I}(z_{2,1})}{{\cal I}(w)}
G^+(z_{1,2},w) \left(G^+(z_1,z_2) + G^-(z_1,z_2)\right) \right].
\eean

\section{Acknowledgements}
It is a pleasure to thank Kaj Roland and Luciano Girardello for
useful discussions.

This work is partially supported by the European Commission TMR programme
ERBFMRX-CT96-0045 in which A.P.\ and M.P.\ are associated to the Milan 
University.


\begin{thebibliography}{999}
%
\bibitem{Kap} V.S. Kaplunovsky, Nucl.Phys. {\bf B307} (1988) 145 
              [hep-th/9205070].
\bibitem{BK} Z. Bern and D. Kosower, Nucl.Phys. {\bf B379} (1992) 451.
\bibitem{DiV} P. Di Vecchia, A. Lerda, L. Magnea, R. Marotta and R. Russo,
              Nucl.Phys. {\bf B469} (1996) 235 [hep-th/9601143].
\bibitem{At} J.J. Atick, L.J. Dixon and A. Sen, Nucl.Phys. {\bf B292} (1987)
             109;\hfill\break
             J.J. Atick and A. Sen, Phys.Lett. {\bf 186B} (1987) 339,
             Nucl.Phys. {\bf B286} (1987) 189 and {\bf B293} (1987) 317.
\bibitem{PR1} A. Pasquinucci and K. Roland, Nucl.Phys. {\bf B440} (1995) 441
              [hep-th/9411015].
\bibitem{KLT} H. Kawai, D.C. Lewellen and S.H.H.Tye, Nucl.Phys.
              {\bf B288} (1987) 1.
\bibitem{Anto} I. Antoniadis, C. Bachas, C. Kounnas and P. Windey,
               Phys.Lett. {\bf 171B} (1986) 51; \\
               I. Antoniadis, C. Bachas and C. Kounnas, 
               Nucl.Phys. {\bf B289} (1987) 87;\\
               I. Antoniadis and C. Bachas, Nucl.Phys. {\bf B298} (1988) 586.
\bibitem{Bluhm} R. Bluhm, L. Dolan and P. Goddard, Nucl.Phys. {\bf B309}
               (1988) 330.
\bibitem{PR2} A. Pasquinucci and K. Roland, Phys.Lett. {\bf B351} (1995) 131
              [hep-th/9503040].
\bibitem{PR3} A. Pasquinucci and K. Roland, Nucl.Phys. {\bf B457} (1995) 27
              [hep-th/9508135].
\bibitem{FMS} D. Friedan, E. Martinec and S. Shenker, Nucl.Phys.
              {\bf B271} (1986) 93.
\bibitem{Kaj} G. Cristofano, R. Marotta and K. Roland, Nucl.Phys. 
              {\bf B392} (1993) 345.
\bibitem{DPh} For a review, see E. D'Hoker and D.H. Phong, Rev.Mod.Phys. 
              {\bf 60} (1988) 917. 
\bibitem{LeB} M. Le Bellac, ``{\sl Des ph\'enom\`enes critiques aux champs de
              jauge\/}'', InterEditions et Editions du CNRS, Paris, 1988. 
\bibitem{Fey} L.M. Brown and R.P. Feynman, Phys.Rev. {\bf 85} (1952) 231.
\bibitem{Wein} S. Weinberg, ``{\sl Radiative corrections in string theory\/}'',
               Talk given at the APS meeting 1985, in DPF Conf. 1985, p. 850.
\bibitem{Fis} W. Fischler and L. Susskind, Phys.Lett. {\bf B171} (1986) 383,
              and {\bf B173} (1986) 262.
\bibitem{Tsey} A.A. Tseytlin, in ``{\sl Trieste Superstrings 1989\/}'', 
               487, and references therein. 
\bibitem{DP2} K. Aoki, E. D'Hoker and D.H. Phong, Nucl.Phys. {\bf B342} 
              (1990) 149;\hfill\break
              E. D'Hoker and D.H. Phong, Phys.Rev.Lett. {\bf 70} (1993) 3692, 
              Theor.Math.Phys. {\bf 98} (1994) 306 [hep-th/9404128], 
              Nucl.Phys. {\bf B440} (1995) 24 [hep-th/9410152].
\bibitem{Bere} A. Berera, Nucl.Phys. {\bf B411} (1994) 157.
\bibitem{Mont} J.L. Montag and W.I. Weisberger, Nucl.Phys. {\bf B363}
              (1991) 527.  
\bibitem{PR5} A. Pasquinucci and K. Roland, Nucl.Phys. {\bf B485} (1997) 241
              [hep-th/9608022].
\bibitem{Koste} V.A. Kostelecky, O. Lechtenfeld, W. Lerche, S. Samuel and
                S. Watamura, Nucl.Phys. {\bf B288} (1987) 173.
\bibitem{Fay} J. Fay, ``{\sl Theta functions on Riemann Surfaces\/}'', Lecture
              Notes in Mathematics {\bf 352}, Springer-Verlag, Berlin, 1973; 
              \hfill\break D. Mumford, ``{\sl Tata Lectures on Theta 1\/}'', 
              Progress in Mathematics {\bf 28}, Birkh\"auser, Boston, 1983;
              \hfill\break
              J. Fay, Duke Math. J. {\bf 51} (1984) 109; \hfill\break
              J. Fay, Proceedings of Symposia in Pure Mathematics {\bf 49} 
              (1989) 485.


\end{thebibliography}
\end{document}